\documentclass[12pt]{iopart}

\usepackage{hyperref}
\expandafter\let\csname equation*\endcsname\relax
\expandafter\let\csname endequation*\endcsname\relax
\usepackage{amsmath}
\usepackage{balance}
\usepackage{graphicx}
\usepackage{color}
\usepackage{setspace}
\usepackage{comment}
\usepackage{xcolor}
\usepackage{array}
\usepackage{multirow}
\usepackage[justification=raggedright]{caption}
\usepackage{colortbl}

\newcommand{\sech}{{\rm sech}}
\begin{document}

\title{Solitons in media with mixed, high-order dispersion and cubic nonlinearity}

\author{Y. Long Qiang$^{1}$,Tristram J. Alexander$^{1}$,C. Martijn de Sterke$^{1,2}$}

\address{$^{1}$ Institute of Photonics and Optical Science (IPOS), School of Physics, The University of Sydney, NSW 2006, Australia.\\
$^{2}$ The University of Sydney Nano Institute (Sydney Nano), The University of Sydney, NSW 2006, Australia}
\ead{yqia7452@uni.sydney.edu.au}

\vspace{10pt}
\begin{indented}
\item[]\today
\end{indented}

\begin{abstract}
Although most soliton research has traditionally considered dominant quadratic dispersion, the recent discovery of pure-quartic solitons has inspired analysis of soliton solutions with large higher orders of dispersion. Here we present analytic expressions for families of bright soliton solutions at arbitrary dispersion orders and practical methods to obtain the associated dispersion relations. These results provide a framework for considering higher order dispersion solitons and show the potential for further investigation of solitons in higher order dispersion systems. 
\end{abstract}

%
%
%
%
%


\section{Introduction}

Solitons are a striking phenomenon that are unique  to physical systems with nonlinear effects \cite{Scott_Chu_McLaughlin_1973, Polturak_1981, r}. In optics, solitons can occur from the balance of group velocity dispersion and the nonlinear Kerr effect, resulting in pulses which maintain their shape upon propagation. They have been used in the study and development of various optical applications including telecommunications \cite{Nakazawa_1994, Mollenauer_1991, Haus_1996}, supercontinuum generation \cite{Dudley_2006, n} and ultrafast lasers \cite{o, Turitsyn_2012}. 

Traditionally, the majority of bright soliton studies and applications has involved negative quadratic dispersion, corresponding to the lowest order dispersion permitting soliton formation, whereas higher order terms were treated as a nuisance which needed to be suppressed or managed \cite{Elgin_1992, Kodama_1994, Hook_1993, Aceves_1994}. More recently, there has been experimental confirmation of the existence of solitons that rely on the balance of the Kerr effect and pure, negative higher-order even dispersion \cite{Blanco_Redondo_2016, Runge_2020,runge_qiang_alexander_rafat_hudson_blanco_redondo_desterke_2021b,deSterke_2021}. While initial experiments were carried out in a waveguide with pure negative $4^{\rm th}$ order dispersion, the development of a fibre laser with programmable dispersion has allowed demonstration of solitons with pure negative high-order dispersion of up to $10^{\rm th}$ order \cite{Runge_2020, runge_qiang_alexander_rafat_hudson_blanco_redondo_desterke_2021b,deSterke_2021}. 

Theoretical studies carried out in parallel have aimed to find analytic solutions in the presence of high-order dispersion. While no analytic solutions are known for pure, negative, even dispersion of order $4$ or higher and a Kerr nonlinearity, some solutions in the presence of mixed orders of dispersion have been reported. Karlsson and H\"o\"ok reported a stable stationary exact solution in the presence of both $2^{\rm nd}$ and $4^{\rm th}$ order dispersion and a pure Kerr nonlinearity in the form of a hyperbolic secant squared \cite{Karlsson_1994}, subsequently confirmed by Pich\'e et al.  \cite{Piche_1996}. More recently, we reported a solution in the form of a stationary hyperbolic secant cubed in the presence of $2^{\rm nd}$, $4^{\rm th}$ and $6^{\rm th}$ orders of dispersion \cite{Qiang_2022}. In 2020 Kudryashov reported a more general set of such solutions, both stationary and moving, in the form of hyperbolic secant solutions up to power $6$, as well as solutions consisting of the sum of two terms, each of which is an integer power of a hyperbolic secant \cite{Kudryashov_2020}. Kudryashov's method requires the solution of a set of algebraic equations, but these become increasingly complicated and unwieldy as the powers and the number of terms, and simultaneously the number of dispersion orders, increase. As a consequence some of the underlying properties of the solutions were not apparent. While Kudryashov considered a pure Kerr nonlinearity, in subsequent work this was generalized to systems with more general nonlinearities \cite{hosseini2021, kudryashov_2021a, kudryashov_2021b,arnous2022}.


Here we develop a systematic method for finding stationary analytic solutions in the form of superpositions of an arbitrary number of hyperbolic secant functions of arbitrary integer power, i.e. superpositions of $\rm sech^p(\alpha\tau)$, where $\tau$ is the time in a co-moving frame, $\alpha$ is a width parameter and $p$ is a positive integer. Since our approach is a systematic one it is straightforward to generalise the solutions reported by Kudryashov \cite{Kudryashov_2020}. We fix the form of the solution, and find the associated dispersion relation. We show that for solutions consisting of a single term, the associated dispersion relation is associated with a polynomial with even power terms and $2p$ roots at integer values of $\alpha$, immediately giving the associated dispersion relation. Since this method cannot be applied to solutions that are superpositions of integer powers of hyperbolic secants, we develop another systematic method in which we cast the problem in the language of linear algebra, and exploit the linear dependence of the equations we obtain. In this way we straightforwardly generate the relevant algebraic equations, which can then be solved by standard methods and find solitons not reported by Kudryashov \cite{Kudryashov_2020}. We also show that the poles of the spectra of the solutions lie on the imaginary axis at integer or half-integer multiples of $i\alpha$. 

The outline of this paper is as follows. In Section~\ref{sec:model}, we introduce the form of the nonlinear Schr\"odinger equation to which we find analytic soliton solutions. In Section~\ref{sec:analytic}, we consider solutions consisting of a single term, reporting the form of these solutions for arbitrary powers and outline how to find the associated dispersion relation. In Section~\ref{sec:analyticpp2}, we consider solutions consisting of a superposition of the powers of two hyperbolic secant terms. We also present the systematic method involving linear algebra which we use to find the associated dispersion relation. In Sec.~\ref{sec:allp}, we generalise the analytic solution for a superposition to an arbitrary number of terms, and the linear algebra method to find the associated dispersion relation. In each section, we also show the unique pole structure in the spectral domain for each of our solutions. 

\section{Generalised Nonlinear Schr\"odinger Equation}\label{sec:model}
We consider a variety of analytic soliton solutions in  the presence of high even orders of dispersion. The evolution of these solutions satisfies the generalised nonlinear Schr\"odinger equation
\begin{align}
i\frac{\partial\psi}{\partial z}+\sum_{r =1}^{n}(-1)^{r}\frac{\beta_{2r}}{(2r)!}\frac{d^{2r}\psi}{d\tau^{2r}}+\gamma |\psi|^{2}\psi= 0.
\label{eq:gnls}
\end{align}
where $\psi(z, \tau)$ is the pulse envelope, $z$ the propagation coordinate, $\tau$ the local time, $\gamma$ the nonlinear parameter, which we take to be positive, and $\beta_{n}$ is the highest order dispersion coefficient.
To find stationary solutions, we set $\psi(z, \tau) = u(\tau; \mu)e^{i\mu z}$, so that $u$ satisfies
\begin{align}
-\mu u+\sum_{r = 1}^{n}(-1)^{r}\frac{\beta_{2r}}{(2r)!}\frac{d^{2r}u}{d\tau^{2r}}+\gamma u^{3}= 0.
\label{eq:stationaryeq}
\end{align}
We note in particular, that since we aim to observe bright, pulse-like solutions, we require $\beta_{n}<0$ and $\mu>0$ \cite{Qiang_2022}. 

We briefly consider the linear wave solutions to Eq.~\eqref{eq:stationaryeq} in the limit in which the wave amplitude is small, allowing us to drop the $\gamma u^3$ term. The resulting equation has solutions given as the linear combination of $n$ terms of the form $\exp(\lambda \tau)$ \cite{runge_qiang_alexander_rafat_hudson_blanco_redondo_desterke_2021b}. The $\lambda$ satisfy the algebraic equation
\begin{equation}
-\mu +\sum_{r=1}^{m}(-1)^{r}\frac{\beta_{2r}}{(2r)!}\lambda^{2r}= 0,
\label{eq:character}
\end{equation}
to which we will refer to as the {\sl characteristic polynomial} for each soliton solution.

\section{Analytic Solution: Single term \texorpdfstring{sech$^{p}$}{}}\label{sec:analytic}

Analytic stationary solutions to problems with nonlinearity and dispersion are fairly rare. However, solutions of the type 
\begin{equation}
    u=A~\sech^p(\alpha\tau)
\label{eq:sechgen}
\end{equation}
are known. The value $p=1$ corresponds to conventional nonlinear Schr\"odinger solitons with negative quadratic dispersion \cite{Zakharov_1972,Ablowitz_Segur_1981}. Karlsson and H{\"o}{\"o}k showed that $p=2$ corresponds to a solution for mixed $2^{\rm nd}$ and $4^{\rm th}$ order dispersion \cite{Hook_1993}. Similarly, a solution for $p=3$, in the presence of  $2^{\rm nd}$, $4^{\rm th}$ and $6^{\rm th}$ order dispersion has been found \cite{Kudryashov_2020, Qiang_2022}. An obvious question that arises is whether such analysis can be extended to higher values of $p$, and whether this analysis can be carried out in a systematic way. We address both of these questions in Section~\ref{sec:p=p}. Before doing so, we note that the way we have cast the problem is to assert a solution (of the form~\eqref{eq:sechgen}), and then to find the associated dispersion relation that solves Eq.~\eqref{eq:stationaryeq}. This contrasts with the usual approach, where, given a dispersion relation, the associated solution needs to be found. 

\subsection{\texorpdfstring{Sech$^{3}$}{} solution}\label{sec:p=3}

Here we search for the dispersion coefficients $\beta_{n}$ and $\mu$ for which Eq.~\eqref{eq:stationaryeq} has solutions of the form~\eqref{eq:sechgen} for arbitrary integer $p$. This problem can be solved by using the relation     
\begin{align}
    \frac{d^{2} \sech^{r}(\tau)}{d\tau^{2}}=r^{2} \sech^{r}(\tau)-r(r+1) \sech^{r+2}(\tau).
    \label{eq:2ndde}
\end{align}
Applied recursively, this expression allows even derivatives of hyperbolic secants to be written as a sum of hyperbolic secant terms. For the highest order dispersion term with an $n$ power derivative in Eq.~\eqref{eq:stationaryeq}, we can obtain hyperbolic secant terms of power $r$ up to $r + n$, and the power of the terms which can be generated for all the different dispersion terms are shown in Table~\ref{table:1}. In this table, each row corresponds to a term in Eq~\eqref{eq:stationaryeq}, whereas each column groups terms of the same power. The existence of a stationary analytic solution requires the entries in each column to cancel. 

\begin{table}[h!]
\centering
\begin{center}
\begin{tabular}{|>{\centering\arraybackslash}p{11mm}|>{\centering\arraybackslash}p{11mm}|>{\centering\arraybackslash}p{1.1cm}|>{\centering\arraybackslash}p{1.1cm}|>{\centering\arraybackslash}p{1.1cm}|>{\centering\arraybackslash}p{1.1cm}|>{\centering\arraybackslash}p{1.1cm}|}
 \hline
 & \multicolumn{5}{|c|}{Power of Hyperbolic Secant terms}  \\ & \multicolumn{5}{|c|}{generated by terms in the general NLS} \\
\hline
$\mu$ & $p$ &  & & & \\
\hline
$\beta_2$ & $p$ & $p+2$ & & &\\
\hline
$\beta_4$ & $p$ & $p+2$ & $p+4$ & & \\
 \hline
 $\vdots$ & $\vdots$ & $\vdots$ & $\vdots$ & $\ddots$ & \\
\hline
$\beta_n$ & $p$ & $p+2$ & $p+4$ & $\cdots$ & $p+n$\\
\hline
$\gamma$ & &  & & & $3p$ \\
\hline
\end{tabular}

\caption{Rows show the powers of hyperbolic secant terms generated by each term through Eq.~\eqref{eq:sechgen}. Columns group the hyperbolic secant terms of equal power}
\label{table:1}
\end{center}
\end{table}

We first note that the nonlinear term needs to cancel the highest power term generated by the dispersion, i.e., $p+n=3p$, or $n=2p$ (see last column of Table~\ref{table:1}). This enforces the relationship between $n$ and $p$ in the general case. Choosing the coefficients such that the entries in this column cancel gives a relation between $\beta_{2p}$, $\gamma A^2$, and $\alpha$. The remaining dispersion coefficients and $\mu$ can be found by solving $p$ simultaneous linear equations. These equations can be written as a triangular matrix (see Table~\ref{table:1}) so they are straightforward to solve. For $p=3$, the parameters are given in Table~\ref{table:termsfirst}.

\begin{table}[h!]
\renewcommand{\arraystretch}{1.5}
\centering
\begin{center}
\begin{tabular}{|>{\centering\arraybackslash}p{11mm}|>{\centering\arraybackslash}p{30mm}|}
\hline
$\gamma A^2$ & $28|\beta_{6}|\alpha^{6}$\\
\hline
$\alpha^2$ & $-\frac{30}{83}\frac{\beta_{4}}{|\beta_{6}|}$\\
\hline
$\mu$ & $\frac{245|\beta_{6}|}{16}\alpha^{6}$ \\
\hline
$\beta_2$ & $\frac{1891|\beta_{6}|}{360}\alpha^{4}$\\
\hline
\end{tabular}

\caption{Exact expressions for the unknowns of Eq.~\eqref{eq:sechgen} for $p=3$}
\label{table:termsfirst}
\end{center}
\end{table}
This approach and the results obtained here are consistent with results found earlier \cite{Kudryashov_2020, Qiang_2022}. Clearly, the number of simultaneous equations that need to be solved increases with $p$. Whereas this becomes increasingly tedious as $p$ increases, below we present an alternative approach that makes this unnecessary. 

\subsection{Sech$^{p}$ solution}\label{sec:p=p}%
To extend our method to arbitrary dispersion order, we consider the last column of Table~\ref{table:1}. As discussed, it represents the cancellation of the nonlinear term in Eq.~\eqref{eq:stationaryeq} and the highest power hyperbolic secant from the dispersion. This term can be generated by recursively applying Eq.~\eqref{eq:2ndde}. Using this it is found that 
\begin{align}
    \gamma A^2={\beta_{2p}\over (2p)!} p(p+1)\ldots(3p-1)\alpha^{2p}={1\over3}{(3p)!\over p!(2p)!}\beta_{2p}\alpha^{2p}.
    \label{eq:general amplitude}
\end{align}
The amplitude in Eq.~\eqref{eq:sechgen} is thus expressed in the dispersion order and $\alpha$. For $p=2$ and $p=3$, this generalised expression matches both the Karlsson and H{\"o}{\"o}k solution and known $6^{\rm th}$ order analytic solution, given in row 1 of Table~\ref{table:termsfirst}, respectively \cite{Qiang_2022}. 

To analyse the linear components for arbitrary order, we note that characteristic polynomials of the Karlsson and H{\"o}{\"o}k solution and known $6^{\rm th}$ order solution, have a distinct root structure. We find the roots to be positioned at $\pm 2\alpha,\pm 4\alpha$, and $\pm 3\alpha,\pm 5\alpha,\pm 7\alpha$, respectively, on the real axis. The roots are thus positioned at integer values of $\alpha$. This can be understood as follows: the asymptotic expansion of Eq.~\eqref{eq:sechgen} is a superposition of terms of the type $e^{\pm p\alpha\tau}$, $e^{\pm (p+2)\alpha\tau}$, $e^{\pm (p+4)\alpha\tau}$, etc. Since $e^{\pm 3p\alpha\tau}$ is the third harmonic of the first term, it can be generated nonlinearly, and similarly for higher powers. Terms between powers $p$ and $3p-2$ cannot be generated nonlinearly, and must be associated with the solutions of the characteristic polynomial. 

We can thus construct the characteristic polynomial by recognising that it must have $2p$ roots at $\pm p \alpha, \pm (p+2) \alpha, \ldots ,\pm (3p-2) \alpha$. It must thus be of the form
\begin{align}
    \left(\!{\lambda^2}\! -(\alpha p)^2\right)\!
    \left(\!{\lambda^2}\! -(\alpha(p+2))^2\right)\! \ldots\! \left(\!{\lambda^2}\! -(\alpha(3p-2))^2\right)\!=0.
    \label{sechppolylambda}
\end{align}
apart from a multiplicative constant. 

Working backwards, we replace $\lambda$ by $\partial/\partial t$ and find 
\begin{align}
    \left(\!{\partial^2\over\partial t^2}\! -(\alpha p^2)\right)\!
    \left(\!{\partial^2\over\partial t^2}\! -(\alpha(p+2))^2\right)\! \ldots\! \left(\!{\partial^2\over\partial t^2}\! -(\alpha(3p-2))^2\right)\!u=0.
    \label{sechppoly}
\end{align}
This expression is equivalent to the linear part of the stationary nonlinear Schr\"odinger equation Eq.~\eqref{eq:stationaryeq}. Multiplying this by $|\beta_{2p}|/(2p)!$ and comparing with the linear terms of Eq.~\eqref{eq:stationaryeq} allows us to obtain the parameters, and linear dispersion coefficients for the general solution Eq.~\eqref{eq:sechgen}. This produces expressions which are  much easier to find than using the method outlined in Section~\ref{sec:p=3}. 

We now use this approach to obtain a general expression for $\alpha$. We take the two highest order equations generated using the method described above, which provides equations for $\beta_{2p}$ and $\beta_{2p-2}$ in terms of $\alpha$. Solving these simultaneously, gives
\begin{align}
    \alpha^{2}=\frac{(2p-1)}{\frac{13}{24}(2p)^{2}-2p+\frac{1}{3}}\frac{\beta_{2p-2}}{\beta_{2p}}.
    \label{eq:alphagen}
\end{align}
This expression, in conjunction with the amplitude from Eq.~\eqref{eq:general amplitude} provides expressions for the unknowns in Eq.~\eqref{eq:sechgen} at arbitrary order. Note that Eq.~\eqref{eq:alphagen} is consistent with row~2 of Table~\ref{table:termsfirst} for $p=3$, as required. 

With $A$ and $\alpha$ known, we require expressions for the linear dispersion relation and the nonlinear phase shift. We can match the prefactors between the expanded polynomial Eq.~\eqref{sechppoly} and the linear part of the stationary NLS (Eq.~\eqref{eq:stationaryeq}), since they are equivalent expressions, which give us the linear dispersion terms and nonlinear phase term, as functions of $\alpha$, $\beta_{2p}$ and $\beta_{2p-2}$. Thus, by choosing values, the full linear dispersion relation and nonlinear phase shift become fully defined. For $p=3$, the polynomial is
\begin{align}
    \frac{\partial^{6}u}{\partial\tau^{6}}-83\alpha^{2}\frac{\partial^{4}u}{\partial\tau^{4}}+1891\alpha^{4}\frac{\partial^{2}u}{\partial\tau^{2}}-11025\alpha^{6}u=0.
    \label{eq:retrieved6th}
\end{align}
Drawing an equivalence with the linear part of the associated stationary NLS, results in the same parameters obtained as in Table~\ref{table:termsfirst}.

An increase in dispersion order corresponds to an increase in the number of constraints (number of columns in Table~\ref{table:1}), whereas the degrees of freedom ($A$, $\alpha$, $\mu$) remains equal to $3$. Thus, for every increase in constraints, an additional dispersion term is required. In fact, apart from $\beta_{2p}$ and $\beta_{2p-2}$, all dispersion terms are prescribed. Geometrically, the dimension of the corresponding phase space increases with dispersion order, whereas the dimension of the solution space for these exact analytic solutions remains the same. For example, the solutions for $p = 3$ correspond to a single curve in a $3$-dimensional space when $\beta_6$ and $\gamma$ are fixed \cite{Qiang_2022}. 

\subsection{Properties of the solutions}\label{sec:prop}
It is straightforward to show that a pulse with an intensity of ${\rm sech}^{2p}(\alpha\tau)$ has a full-width at half-maximum (FWHM)
\begin{align}
    w = \frac{2}{\alpha}\ln\left(2^{\frac{1}{2p}}+\sqrt{2^{\frac{1}{p}}-1}\right)\approx\frac{2}{\alpha}\sqrt{\frac{\ln 2}{p}},
\label{eq:fwhmsech4}
\end{align}
where the last result holds when $p\gg1$. In this limit $\alpha^2$ decreases proportionally with $p$, which cancels with the narrowing of ${\rm sech}^{2p}(x)$ as $p$ increases; as a consequence, the FWHM takes a limiting value. Using Eq.~\eqref{eq:alphagen} we find, as $p\rightarrow\infty$
\begin{align}
    w \approx \sqrt{\frac{13\ln 2}{3}\frac{|\beta_{2p}|}{\beta_{2p-2}}}.
\label{eq:fwhmsech4ap}
\end{align}

\begin{figure}[hbt]
\centering
\hspace*{0cm}\includegraphics[width=120mm,clip = true]{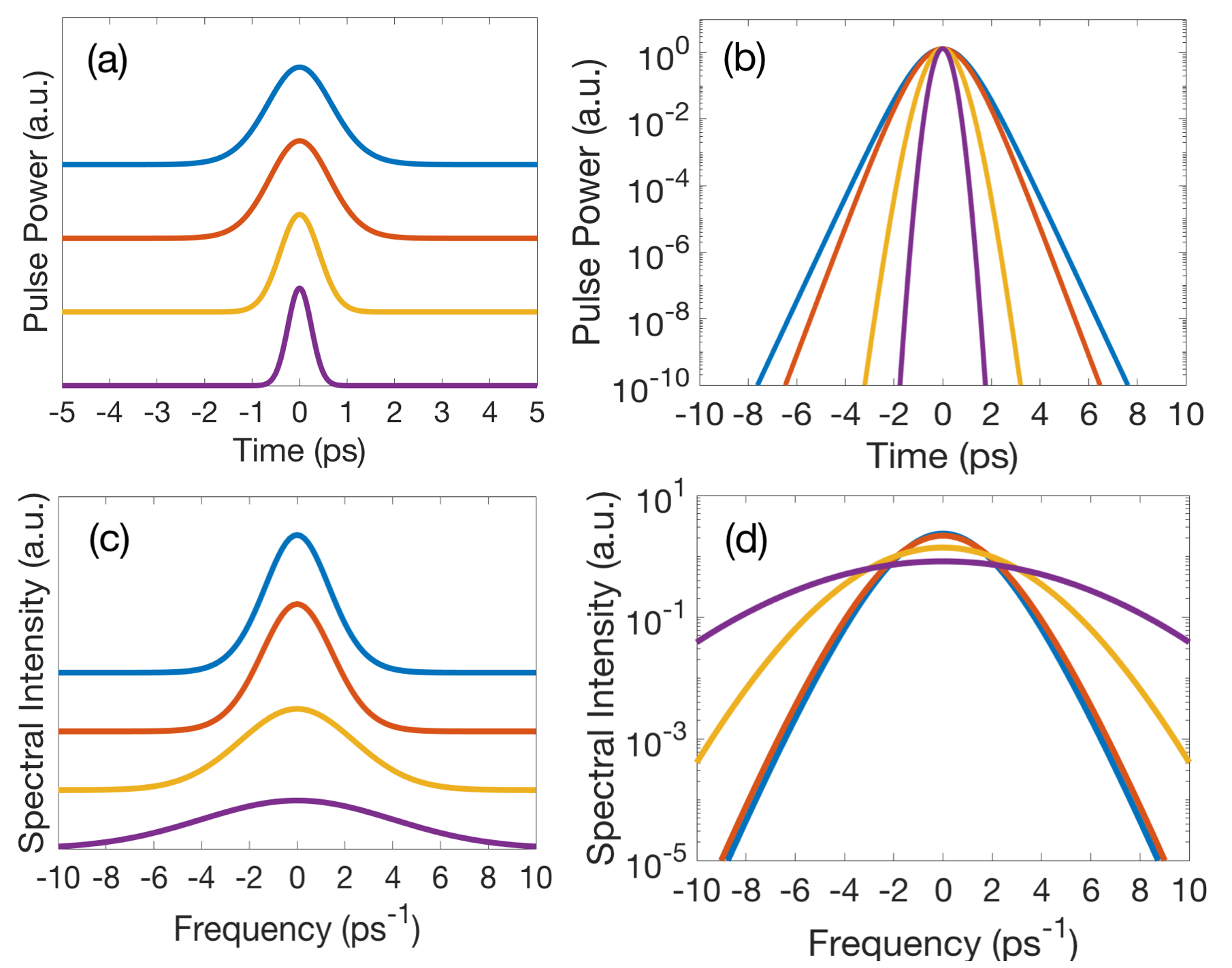}
\vskip 0mm
\caption{\label{fig:exact}
Power of solutions of the form~\eqref{eq:sechgen}, with peak power $P=1.3222~{\rm W}$, $\alpha= 0.6012~{\rm ps^{-1}}$ and $\gamma=1~ {\rm W}^{-1} {\rm mm}^{-1}$ kept constant, for $p=3$ (blue: $\beta_{4}=-1\ {\rm ps^{4}~ mm}^{-1}$, $\beta_{6}=-1\ {\rm ps^{6}~ mm}^{-1}$, ${\rm FWHM} = 1.63~{\rm ps}$), $p=4$ (red: $\beta_{6}=-1.1985\ {\rm ps^{6}~ mm}^{-1}$, $\beta_{8}=-0.8597\ {\rm ps^{8}~ mm}^{-1}$, ${\rm FWHM} = 1.5152~{\rm ps}$), $p=15$ (yellow: $\beta_{28}=-3.0689\ {\rm ps^{28}~ mm}^{-1}$, $\beta_{30}=-0.5378\ {\rm ps^{30}~ mm}^{-1}$, ${\rm FWHM} = 0.9789~{\rm ps}$), $p=50$ (purple: $\beta_{98}=-8.7665\ {\rm ps^{98}~ mm}^{-1}$, $\beta_{100}=-0.4516\ {\rm ps^{100}~ mm}^{-1}$, ${\rm FWHM} = 0.5835~{\rm ps}$) on (a) a linear scale, and (b) a log scale. Same pulses versus frequency on (c) linear scale, and (d) log scale. The solutions on the linear scale have been shifted for clarity.
}
\vskip-1mm
\end{figure}
Figures~\ref{fig:exact}(a) and (b) show the analytic solutions (Eq.~\eqref{eq:sechgen}) with the same peak power $P=1.322~{\rm W}$ and same width parameter $\alpha=0.6012~{\rm ps^{-1}}$, for $p=3,4,15,50$, on linear and logarithmic scales respectively. The figures show that the tails of the pulses steepen as $p$ increases, corresponding to the increasing strength of the dominant term ($e^{\pm p\alpha\tau}$) of the asymptotic expansion of Eq.~\eqref{eq:sechgen}. The corresponding plots in the frequency domain in Figs.~\ref{fig:exact}(c) and (d) show an opposite effect, with steeper solutions in the temporal domain corresponding to broader solutions in the spectral domain, as expected.

\begin{figure}[hbt]
\centering
\hspace*{0cm}\includegraphics[width=120mm,clip = true]{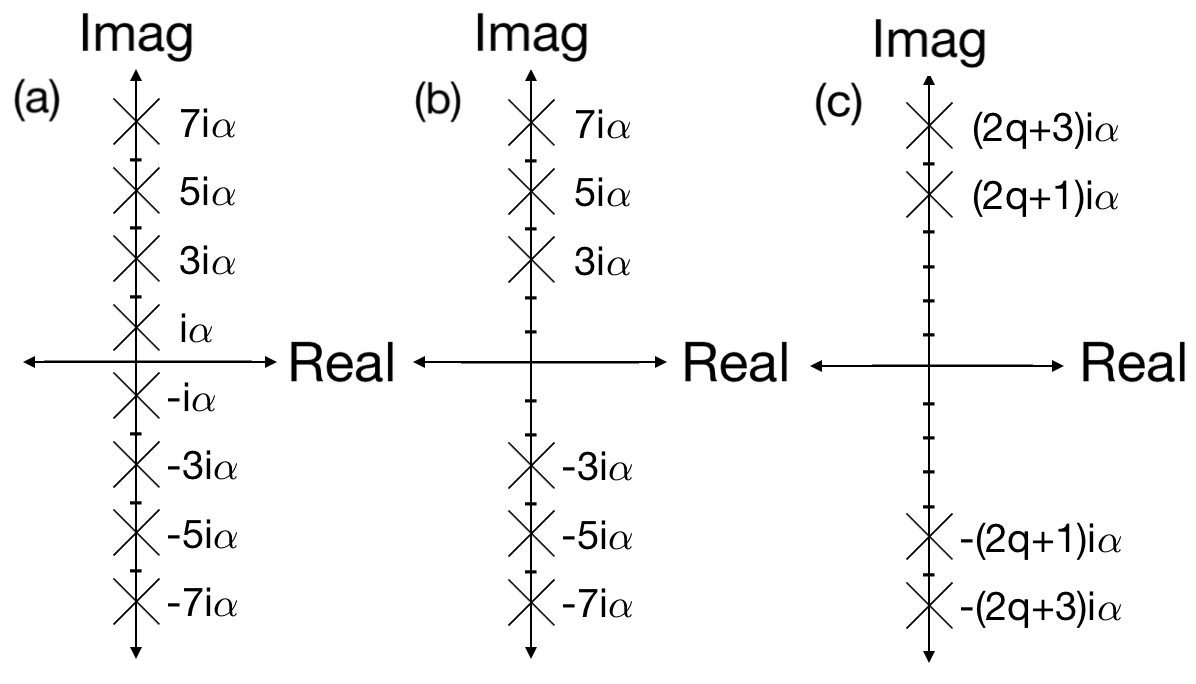}
\vskip 0mm
\caption{\label{fig:infinite}
The positions of the poles for the Fourier Transform for (a) ${\rm sech}(\alpha\tau)$, (b) ${\rm sech^{3}}(\alpha\tau)$ and (c) ${\rm sech^{2q+1}}(\alpha\tau)$.}
\vskip-1mm
\end{figure}

We find that considering the solutions in the spectral domain reveals intriguing properties, and  start by considering in this section the Fourier transform for a single term analytic solution. The Fourier transform of a  hyperbolic secant is \cite{erdelyi_bateman_1954}
\begin{align}
    {\rm FT}\left({\rm sech}(\alpha\tau)\right) = \frac{\pi}{\alpha}{\rm sech}\left(\frac{\pi \omega}{2\alpha}\right),
    \label{eq:FT1}
\end{align}
and is thus also a hyperbolic secant. It has simple poles that are evenly spaced infinitely along the imaginary axis; for our particular function, poles are found at $\omega = \pm i\alpha, \pm 3i\alpha, \pm 5i\alpha, ...$, as indicated in Fig.~\ref{fig:infinite}(a). 

The general analytic form of the Fourier transform for odd powers of the hyperbolic secant is known \cite{erdelyi_bateman_1954}, and can also be shown by induction to be 
\begin{align}
    {\rm FT}\left({\rm sech^{2q+1}}(\alpha\tau)\right) 
    =\frac{4^{q}\pi}{(2q)!\alpha}{\rm sech}\left(\frac{\pi \omega}{2\alpha}\right)\prod^{q}_{r=1}\frac{1}{4\alpha^2}\left(\omega^2+\alpha^2\left(2r-1\right)^2\right).
    \label{eq:FT2}
\end{align}
The resultant function is the product of Eq.~\eqref{eq:FT1} and a polynomial of order $q^2$. This polynomial has roots at $\omega = \pm i\alpha, ..., \pm (2q-1)i\alpha, \pm (2q+1)i\alpha$. These roots supplant the simple poles of the hyperbolic secant function present at these points. The position of the poles for the Fourier transform of a cubic hyperbolic secant ($q=1$), and the general hyperbolic secant with odd powers are plotted in Fig.~\ref{fig:infinite}(b) and (c) respectively, showing the decreasing number of poles as they are supplanted by the increasing roots. Thus, for every additional dispersion order, the number of poles decreases by two, at frequencies of increasing magnitude. 

An analogous argument can be made for the Fourier transform of even power hyperbolic secants \cite{erdelyi_bateman_1954}. Their Fourier transforms have the same properties except that the hyperbolic secant is replaced by a hyperbolic cosecant multiplied by an odd power polynomial. This solution retains similar properties to the odd order case, where the positions of the poles are evenly spaced along the imaginary axis but at multiples of $2i\alpha$ instead ($\omega = \pm 2i\alpha, \pm 4i\alpha, \pm 6i\alpha, ...$). Thus the same principles apply for even power, with simple poles supplanted by roots at the same frequency. This unique property of these analytic solutions in the spectral domain becomes increasingly useful as we explore additional analytic solutions in Sec.~\ref{sec:analyticpp2} and \ref{sec:allp}.


\section{Analytic Solution: Two terms \texorpdfstring{sech$^{p}$+sech$^{p-2}$}{}}\label{sec:analyticpp2}



Here we consider a solution composed of the sum of two hyperbolic secants, an analytic solution in the form 
\begin{align}
    u = A_1\ {\rm sech}^{p-2}(\alpha\tau)+A_2\ {\rm sech}^{p}(\alpha\tau)
    \label{eq:pp_2}
\end{align}
for $p\geq4$. The condition to satisfy the NLS remains $n = 2p$, for the highest power dispersion term and nonlinear term to cancel.The key to finding such solutions in a systematic way is to use the linear dependence of the equations corresponding to the terms at powers $p-2$ and $p$. Table~\ref{table:2} is the analogue of Table~\ref{table:1} for the two term solutions~\eqref{eq:pp_2}. The two columns corresponding to powers $p-2$ and $p$ both have hyperbolic secant terms generated by each of the linear terms in Eq.~\eqref{eq:stationaryeq}, enforcing a linear dependence between the equations at these two powers. This leads to a relationship between the two amplitudes in Eq.~\eqref{eq:pp_2}, and ultimately allows us to separate the problem of finding the solution to Eq.~\eqref{eq:pp_2} from the problem of finding the associated dispersion relation. As we show in Sec.~\ref{sec:p-2p}, we can then find the parameters in Eq.~\eqref{eq:pp_2} for arbitrary order.  We then separately develop a method, which differs from that in  Sec.~\ref{sec:analytic}, to find the dispersion relation. 


\begin{table}[h!]
\centering
\begin{center}
\begin{tabular}{|>{\centering\arraybackslash}p{7mm}|>{\centering\arraybackslash\columncolor[RGB]{102, 204, 255}}p{7.5mm}|>{\centering\arraybackslash\columncolor[RGB]{153, 255, 204}}p{0.75cm}|>{\centering\arraybackslash\columncolor[RGB]{255, 255, 153}}p{0.75cm}|>{\centering\arraybackslash\columncolor[gray]{1}}p{0.75cm}|>{\centering\arraybackslash\columncolor[RGB]{255, 153, 255}}p{1.4cm}|>{\centering\arraybackslash\columncolor[RGB]{255, 153, 51}}p{1.4cm}|>{\centering\arraybackslash\columncolor[RGB]{0, 255, 0}}p{1.4cm}|>{\centering\arraybackslash\columncolor[RGB]{0, 153, 255}}p{0.95cm}|}
 \hline
 & \multicolumn{8}{|c|}{Power of Hyperbolic Secant terms } \\ & \multicolumn{8}{|c|}{generated by terms in the general NLS} \\
\hline
$\mu$ & $p-2$ & $p$ & & & & & &\\
\hline
$\beta_2$ & $p-2$ & $p$ & $p+2$ & & & & & \\
\hline
$\beta_4$ & $p-2$ & $p$ & $p+2$ & $p+4$ & & & &\\
 \hline
 $\vdots$ & $\vdots$ & $\vdots$ & $\vdots$ & $\vdots$ & $\ddots$ &$\ddots$ & $\ddots$&\\
\hline
$\beta_n$ & $p-2$ & $p$ & $p+2$ & $\cdots$ & $p+n-6$ & $p+n-4$& $p+n-2$& $p+n$\\
\hline
$\gamma$ & &  & & & $3p-6$ & $3p-4$&$3p-2$&$3p$\\
\hline
\end{tabular}
\end{center}
\caption{As Table~\ref{table:1} but for substituting Eq. \eqref{eq:pp_2} into Eq.~\eqref{eq:stationaryeq}. The colours of the columns correspond to the terms of the same colour in Fig.~\ref{fig:p5ex}.}
\label{table:2}
\end{table}
The reason the method from Sec.~\ref{sec:p=p} for finding the dispersion relation does not apply here is that the substitution of Eq.~\eqref{eq:pp_2} into the Kerr nonlinear term leads to four different hyperbolic secant powers (see the last four columns of Table~\ref{table:2}). 
This increased number of terms due to the nonlinearity prevents the tail roots being easily predictable, since they no longer all follow from linear equations as in Section~\ref{sec:p=p}. As shown by Kudryashov \cite{Kudryashov_2020}, it is possible to simultaneously solve the equations corresponding to the columns of Table~\ref{table:2} \cite{Kudryashov_2020}, however this is increasingly tedious as $p$ increases. We therefore look for an alternative method.

\begin{figure}[hbt]
\centering
\hspace*{0cm}\includegraphics[width=150mm,clip = true]{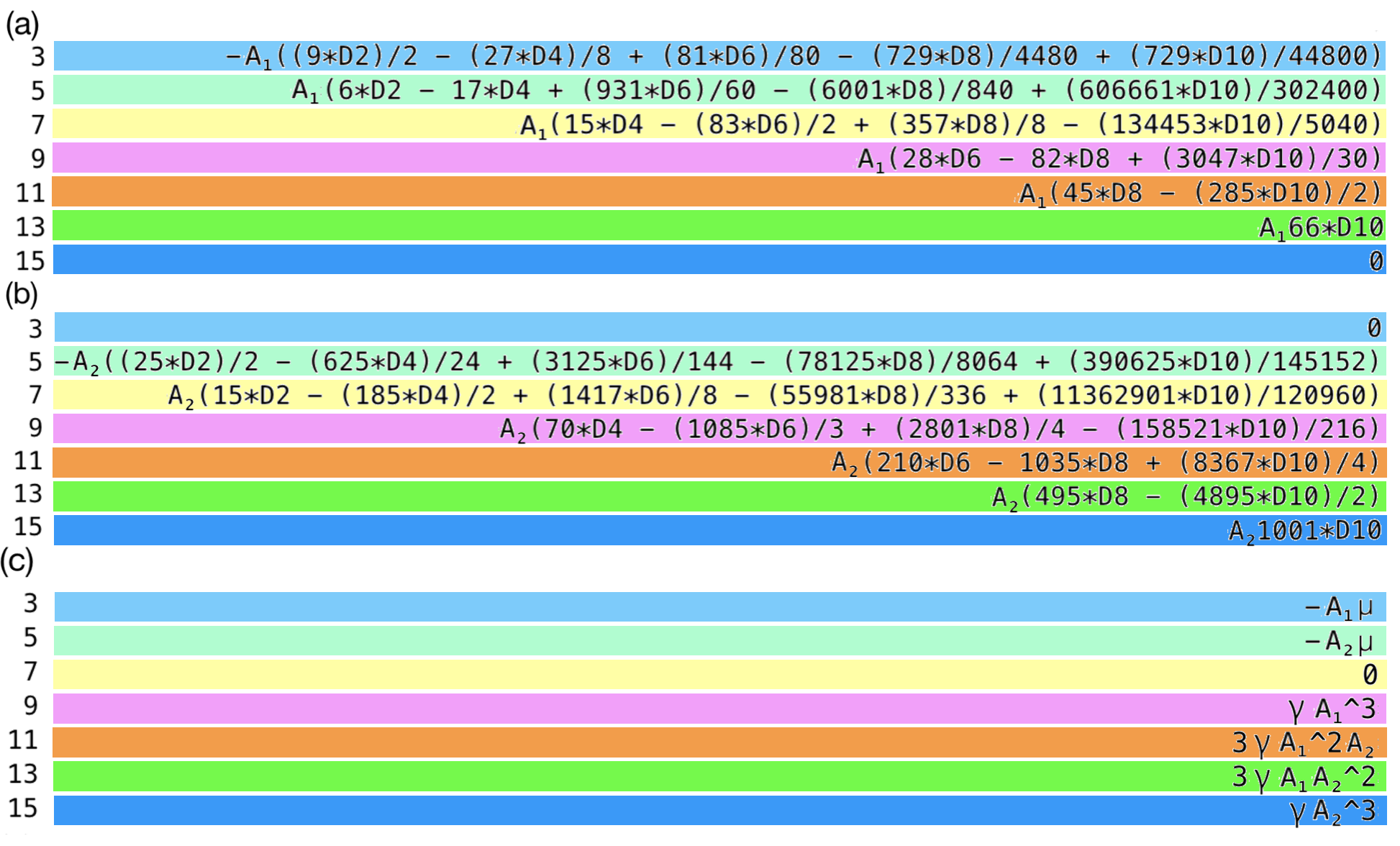}
\vskip 0mm
\caption{\label{fig:p5ex}
Equations generated at each hyperbolic secant power for $p=5$ with  colours corresponding Table~\ref{table:2}. Terms are grouped as follows: (a) from $A\ {\rm sech}^{3}(\alpha\tau)$; (b) from $B\ {\rm sech}^{5}(\alpha\tau)$; (c) nonlinear phase shift $\mu u$ and nonlinear terms $\gamma u^3$. }
\vskip-1mm
\end{figure}

\subsection{Sech$^{3}$+Sech$^{5}$ solution}\label{sec:3-5}

We now take the analytic solution with $p=5$ as an example, giving 
\begin{align}
    u = A_1\ {\rm sech}^{3}(\alpha\tau)+A_2\ {\rm sech}^{5}(\alpha\tau),
    \label{eq:ex35}
\end{align}
and identify a process to obtain the unknowns. Figures~\ref{fig:p5ex}(a)-(c) show all of the terms generated by a substitution of Eq.~\eqref{eq:ex35} into Eq.~\eqref{eq:stationaryeq}, with the colours corresponding to the columns of Table~\ref{table:2} for $p=5$. 

These correspond to the lowest power columns in Table~\ref{table:2} 
and all of the corresponding terms are shown in rows of the same power in Fig.~\ref{fig:p5ex}. These produce proportional equations which we simultaneously solve, giving the ratio $A_2=3A_1/4$, consistent with known results \cite{Kudryashov_2020}. We then substitute this ratio into the unused equations at powers $7, 9, 11, 13, 15$, removing one unknown, and solve the equations for powers $13$ and $15$ simultaneously, giving exact expressions for $\alpha$ and the amplitude $A_1$(see first two rows in Table~\ref{table:terms}). Freely choosing values for $\beta_{10}$, $\beta_8$ and $\gamma$ and substituting these into the remaining unused equations, and simultaneously solving them allows us to find the remaining dispersion terms and the nonlinear phase shift. Exact expressions for these unknowns can be found in the last four rows of Table~\ref{table:terms}. These results mirror existing results and methods by Kudryashov \cite{Kudryashov_2020}, but encounter the same practical issues as expressed in Sec.~\ref{sec:analytic} for higher order dispersion. Similar to the single term case, the number of simultaneous equations to solve increases with dispersion order, and becomes increasingly tedious. We therefore exploit the equations at the two lowest hyperbolic secant powers for the general solution Eq.~\eqref{eq:pp_2}, to develop a different approach.


\begin{table}[h!]
\renewcommand{\arraystretch}{1.5}
\centering
\begin{center}
\begin{tabular}{|>{\centering\arraybackslash}p{11mm}|>{\centering\arraybackslash}p{30mm}|}
\hline
$\alpha^2$ & $\frac{90}{1157}\frac{\beta_{8}}{\beta_{10}}$\\
\hline
$\gamma A_1^2$ & $-\frac{16016}{9}\alpha^{10}\beta_{10}$\\
\hline
$\mu$ & $-\frac{97433}{33}\alpha^{10}\beta_{10}$ \\
\hline
$\beta_2$ & $\frac{17938}{21}\alpha^{8}\beta_{10}$\\
\hline
$\beta_4$ & $\frac{9448}{33}\alpha^{6}\beta_{10}$\\
 \hline
$\beta_6$ & $\frac{27379}{360}\alpha^{4}\beta_{10} $\\
\hline
\end{tabular}

\caption{Exact expressions for the unknowns of Eq.~\eqref{eq:ex35}}
\label{table:terms}
\end{center}
\end{table}
\subsection{Sech$^{p-2}$+Sech$^{p}$ solution}\label{sec:p-2p}
To generalise solutions in the form of Eq.~\eqref{eq:pp_2} to arbitrary consecutive powers, we start with the first two columns of Table~\ref{table:2}. 
For Eq.~\eqref{eq:pp_2} to be a solution, it is necessary that the corresponding equations, can be described as linearly dependent vectors with the same basis. 
Thus the entries in each row must be proportional, such that the prefactor of each ${\rm sech}^{p-2}$ term is linearly proportional to each corresponding ${\rm sech}^{p}$ term. As an example, the ratio of the prefactors for $\mu$ ($A_1$/$A_2$) must be equal to the ratio of the prefactors for $\beta_2$ ($A_1(p-2)^2$/($A_2p^2-A_1(p-2)(p-1)$). Solving this equivalence gives 
\begin{align}
A_1 = \frac{4}{p-2}A_2,
\label{eq:AB}
\end{align}
reducing the number of unknowns. We note that the same ratio is found from any of the other dispersion terms. 

We then consider the last two columns of Table~\ref{table:2}, corresponding to terms at powers $3p-2$ and $3p$. We substitute Eq.~\eqref{eq:AB} into these equations and simplify using some series analysis and factorisation. Solving these simultaneously gives 
\begin{align}
\alpha^2 = &\frac{n(n-1)}{\frac{n}{24}(13n^2-24n+8)-2(n-2)+\frac{6(3n-4)(3n-2)}{n-4}}~\frac{\beta_{n-2}}{\beta_n},\label{eq:alphaexpp2}\\
\gamma A_1^2 &=-\frac{\beta_n\alpha^n}{3} \left(\frac{8}{n-4}\right)^2 \frac{(3n/2)!}{n!(n/2)!},\label{eq:Aexpp2}
\end{align}
where $n=2p$. These results are consistent with the results for $p=5$ in Table~\ref{table:terms}. Choosing values for $\beta_{n}$, $\beta_{n-2}$ and $\gamma$, the remaining dispersion terms and nonlinear phase shift can be found fom the remaining equations. Just like for $p=5$, solving the remaining equations simultaneously gives exact expressions for the dispersion terms and nonlinear phase shift, however we have the same practical issues as existing methods as $p$ increases.

The general expressions for the solution are given in Eq.~\eqref{eq:alphaexpp2} and Eq.~\eqref{eq:Aexpp2}, requires only the amplitude ratio Eq.~\eqref{eq:AB} and two highest power equations to find, thus the majority of the effort then goes into finding the dispersion coefficients using the remaining equations. 
Since our main analytic problem is to solve a set of coupled linear equations, we aim to describe the problem in matrix form and use methods from linear algebra to simplify the process.

This begins by redefining Eq.~\eqref{eq:stationaryeq} using a matrix formalism that incorporates  Eq.~\eqref{eq:2ndde}. This relationship means that every second derivative term enforces two outcomes, either a hyperbolic secant term at the same order with prefactor $r^2$ or a hyperbolic secant term two orders higher with prefactor $-r(r+1)$. For our solution Eq.~\eqref{eq:pp_2}, the lowest power hyperbolic secant is  $p-2$. In matrix form, these derivative terms can be expressed as 
\begin{center}
\begin{align}
D_2 = 
\begin{bmatrix}
(p-2)^2 & 0 & 0&0&\hdots\\
-(p-2)(p-1) & p^2 & 0&0&\hdots\\
0 & -p(p+1) & (p+2)^2& 0&\hdots\\
0 & 0 & -(p+2)(p+3) &(p+4)^4&\hdots\\
\vdots & \vdots & \vdots &\ddots&\ddots\\
\end{bmatrix}
\end{align}
\end{center}
for a $(p+2)\times(p+2)$ matrix. 
The elements $(j,k)$ give the possible prefactors from Eq. \eqref{eq:2ndde}. Column $j$ corresponds to the power of the hyperbolic secant term which Eq.~\eqref{eq:2ndde} acts upon, spanning from $p-2$ to $p+n$ in multiples of two. Row $k$ corresponds to the power of the hyperbolic secant terms resulting from Eq.~\eqref{eq:2ndde}, which span the same range of hyperbolic secant powers as the columns.
This matrix thus describes the action of a $2^{\rm nd}$ order derivative term on hyperbolic secant terms from order $p-2$ to $p+n$, and since the process is linear, we can denote higher order derivative effects as $D_r = D_{2}^{r/2}$.

Next, we describe our analytic solution Eq.~\eqref{eq:pp_2} as a vector 
\begin{align}
    U^T=\begin{bmatrix}
A_1& 
A_2& 
0 &
0 &
\hdots &
    \end{bmatrix}
\end{align}
where $A_1$ and $A_2$ are the amplitudes of our solution from Eq.~\eqref{eq:pp_2}. 
This length of the vector is $p+2$ to match the the matrix $D_r$. We then account for the prefactors of each of the linear terms from Eq.~\eqref{eq:stationaryeq}, including the dispersion terms, the nonlinear phase shift, and the frequency term from Eq.\eqref{eq:pp_2}, expressing it in vector form as
\begin{align}
    B^T=\begin{bmatrix}
-\mu& 
(-1)\frac{\beta_{2}}{2!}\alpha^{2} &
(-1)^2\frac{\beta_{4}}{4!}\alpha^{4} &
\hdots &
(-1)^{p}\frac{\beta_{2p}}{(2p)!}\alpha^{2p}
    \end{bmatrix}
\end{align}
a vector of length $p+1$. Finally, the nonlinear term $\gamma u^3$ can be expressed as
\begin{align}
    N^T=\gamma\begin{bmatrix}
\hdots& 
0&
A_1^3&
3A_1^2A_2& 
3A_1A_2^2 &
A_2^3
    \end{bmatrix},
\end{align}
a vector of length $p+2$, where the final four rows correspond to the rows where nonlinear terms exist, at power $3p-6, 3p-4, 3p-2, 3p$, and where $A_2$ can be eliminated by Eq.~\eqref{eq:AB}.

Now we can express all terms in Eq.~\eqref{eq:stationaryeq} for the solution using matrix notation, as
\begin{align}
\begin{bmatrix}
 U
 & D_2 U & D_4 U & \hdots &  D_n U 
\end{bmatrix}
B= -N.
\label{eq:stationarymatrix}
\end{align}
Here the expression in square brackets represents a $(p+2)\times(p+1)$ matrix, the columns of which consists of the $p+1$ vectors $U,  D_2U, \dots, D_nU$, and represent the $p+2$ linear equations we are aiming to solve. 
The resultant matrix is not square, but this issue is resolved by recalling that the two linear equations at power $p-2$ and $p$ are proportional, allowing us to remove the first row of Eq.~\eqref{eq:stationarymatrix} without losing information. This is true as long as the first two entries of the nonlinear vector $N$ are also zero. This again, is only true for $p\geq 4$. 
This leaves us with a $(p+1)\times(p+1)$ upper triangular square matrix for $p\geq 4$, which can be easily solved as a set of simultaneous linear equations.

Once this set of equations is solved, we can choose values for the two highest order dispersion terms ($\beta_n$ and $\beta_{n-2}$) and the nonlinear parameter ($\gamma$), and we are left with a simple set of equations which describe all the remaining unknowns. This includes $\alpha$ and $A_1$, using the two highest power equations obtained using this matrix process, which is consistent with Eq.~\eqref{eq:alphaexpp2} and Eq.~\eqref{eq:Aexpp2}. This approach using matrices outlines a systematic method to obtain the corresponding unknowns of the solution. Using this method for $p=5$ leads to the same values as in Table~\ref{table:terms}. In \ref{sec:sech13}, we see an example for $p<4$ where a systematic approach cannot be applied since the linear dependence of the two lowest power equations does not hold.

\begin{figure}[hbt]
\centering
\hspace*{0cm}\includegraphics[width=120mm,clip = true]{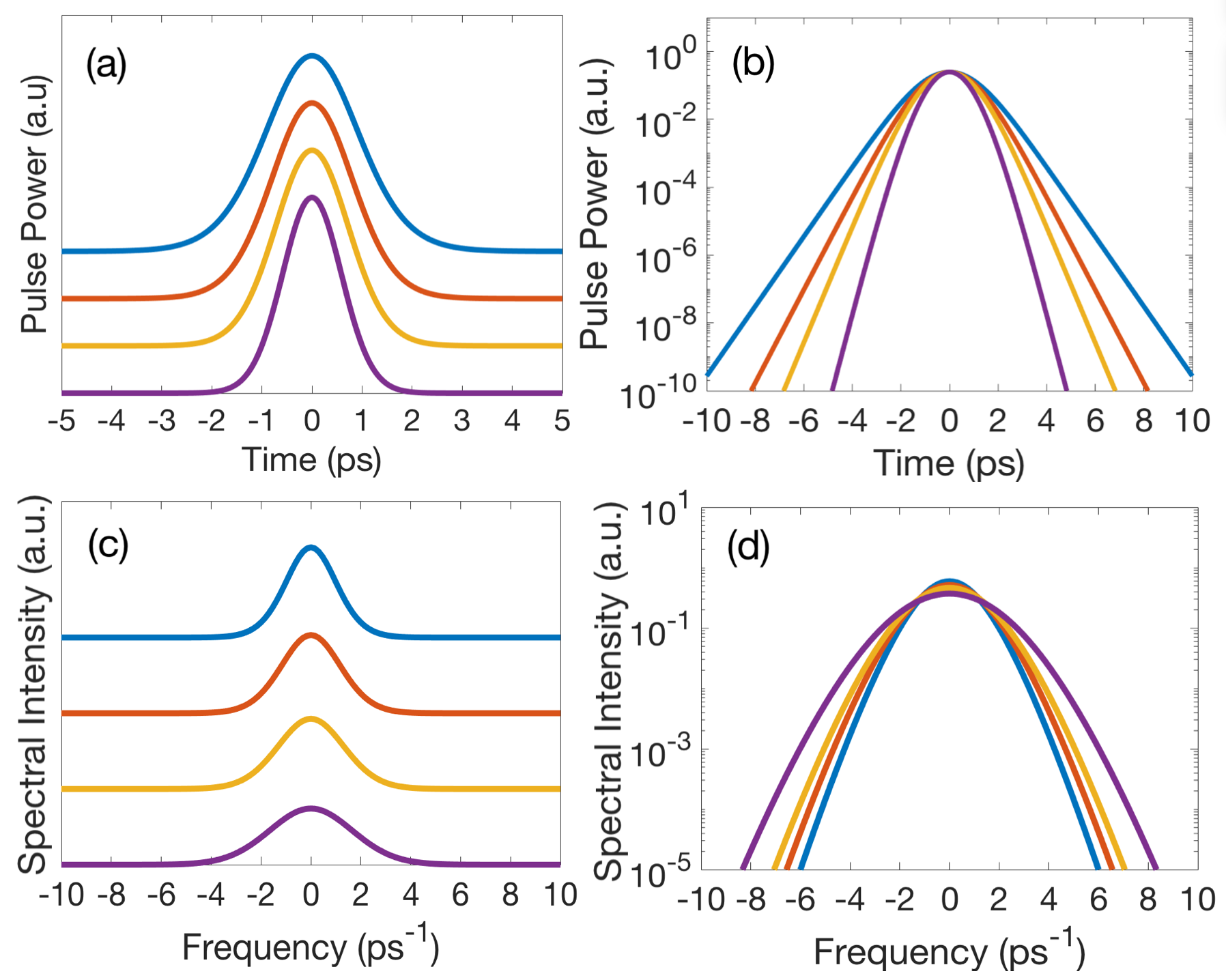}
\vskip 0mm
\caption{Power of analytic solutions~\eqref{eq:pp_2}, with peak power $P = 0.2484 W$, $\alpha=0.3944~{\rm ps^{-1}}$ and $\gamma=1~{\rm W}^{-1} {\rm mm}^{-1}$ kept constant for $p=5$ (blue line: $\beta_{8}=-1\ {\rm ps^{8}~ mm}^{-1}$, $\beta_{10}=-0.5\ {\rm ps^{10}~ mm}^{-1}$, ${\rm FWHM} = 2.1948~{\rm ps}$), $p=6$ (red line: $\beta_{10}=-1.3280\ {\rm ps^{10}~ mm}^{-1}$, $\beta_{12}=-0.7078\ {\rm ps^{12}~ mm}^{-1}$, ${\rm FWHM} = 1.9167~{\rm ps}$), $p=7$ (yellow line: $\beta_{12}=-1.6730\ {\rm ps^{12}~ mm}^{-1}$, $\beta_{14}=-0.8965\ {\rm ps^{14}~ mm}^{-1}$, ${\rm FWHM} = 1.7280~{\rm ps}$), $p=10$ (purple line: $\beta_{18}=-2.7830\ {\rm ps^{18}~ mm}^{-1}$, $\beta_{20}=-1.3271\ {\rm ps^{20}~ mm}^{-1}$, ${\rm FWHM} = 1.3917~{\rm ps}$) on (a) a linear scale, and (b) a logarithmic scale. Associated spectra on (c) a linear scale, and (d) a logarithmic scale. The solutions on the linear scale have been shifted for clarity.}\label{fig:Logsecnn_2}
\vskip-1mm
\end{figure}
\begin{figure}[hbt]
\centering
\hspace*{0cm}\includegraphics[width=100mm,clip = true]{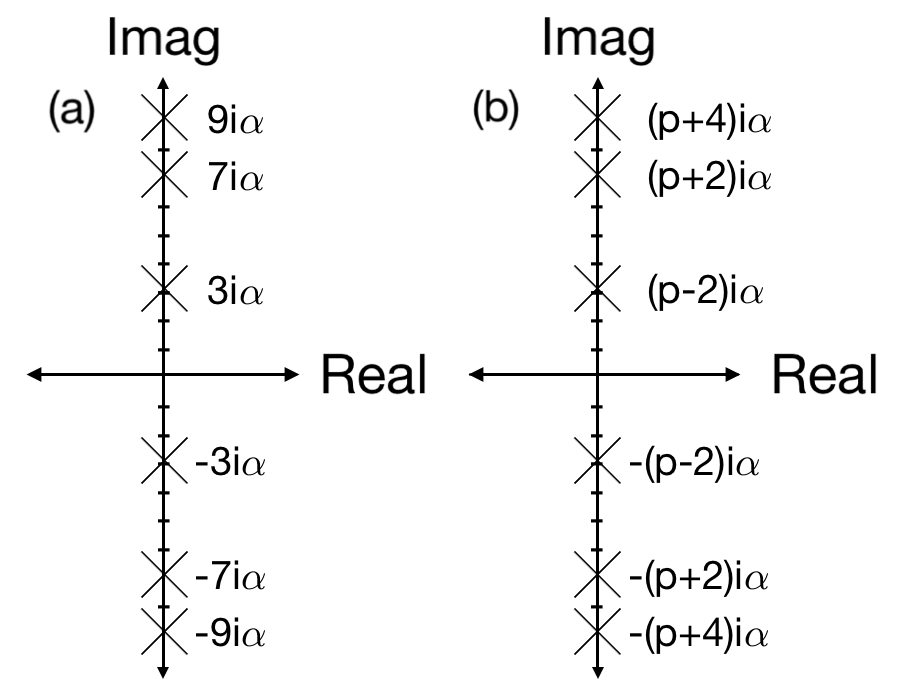}
\vskip 0mm
\caption{\label{fig:ftnn2}
The positions of the poles for the Fourier Transform for (a) $A_1{\rm sech^{3}}(\alpha\tau) +$ $A_2{\rm sech^{5}}(\alpha\tau)$ and (b) $A_1{\rm sech^{p-2}}(\alpha\tau)+A_2{\rm sech^{p}}(\alpha\tau)$.
}
\vskip-1mm
\end{figure}

\subsection{Properties of the solutions}\label{sec:prop2}
We now compare the properties of the solutions~\eqref{eq:pp_2}, to previous solutions Eq.~\eqref{eq:sechgen}. The principles of the tail argument outlined in Sec.~\ref{sec:analytic} remains the same: the smallest real part of the tail solution dominates. For Eq.~\eqref{eq:pp_2}, the smallest real root corresponds to the first term and so we expect the dominant term to be $e^{\pm (p-2)\alpha\tau}$ as $|\tau|\rightarrow\infty$. In contrast, solutions at the same dispersion order for Eq.~\eqref{eq:sechgen} have a dominant term at power $p$. The exponential tails for Eq.~\eqref{eq:pp_2} again become steeper as the dispersion order increases, by the same argument as in Sec.~\ref{sec:p=p}. In Fig.~\ref{fig:Logsecnn_2}, we show solutions~\eqref{eq:pp_2}) for fixed $\alpha$ and $p=5,6,7,10$. Figure.~\ref{fig:Logsecnn_2}(b) shows the increasing steepness as the order increases, with the opposite effect shown in the spectra (Fig.~\ref{fig:Logsecnn_2}(d)).

We now analyze the Fourier transform of Eq.~\eqref{eq:pp_2}, using Eq.~\eqref{eq:AB}. After some rearrangement we find that 
\begin{align}
    {\rm FT}\left(A_1\ {\rm sech^{p-2}}(\alpha\tau)+A_2\ {\rm sech^{p}}(\alpha\tau)\right) 
    =\frac{A_1}{4\alpha^2}\frac{(\omega^2+\alpha^2p^2)}{(p-1)}{\rm FT}({\rm sech^{p-2}}(\alpha\tau)),
    \label{eq:FTtwoorders}
\end{align}
from which we can immediately infer the pole structure. Recall from Sec.~\ref{sec:p=p} that the Fourier transform of a hyperbolic secant of order $p-2$ has simple poles at $\omega = \pm (p-2)i\alpha, \pm pi\alpha, \pm (p+2)i\alpha, \ldots$. However, the simple poles at $\omega = \pm pi\alpha$ are supplanted by the root at the same point which we observe in the RHS of Eq.~\eqref{eq:FTtwoorders}. Thus the ratio Eq.~\eqref{eq:AB} effectively removes one of the simple poles (in fact we find this ratio can be found by enforcing the condition that the simple pole is supplanted). This changes the pole structure, as shown in Fig.~\ref{fig:ftnn2}. 


\section{Analytic solution: Multiple terms} \label{sec:allp}

With the family of analytic solutions established for a sum of two hyperbolic secant terms, we consider possible families, for an arbitrary number of terms, of the form 
\begin{align}
    u = A_1\ {\rm sech}^{p-v}(\alpha\tau)+A_2\ {\rm sech}^{p-v+2}(\alpha\tau)+...+A_{y}\ {\rm sech}^{p}(\alpha\tau)
    \label{eq:ppgen}
\end{align}
for $v$ even, with $p-v$ defining the lowest order term, and the total number of terms $y = v/2+1$. 
Table \ref{table:3} is similar to Tables~\ref{table:1} and~\ref{table:2}, showing the hyperbolic secant terms generated from Eq.~\eqref{eq:ppgen}. We can use the same matrix method developed in Sec.~\ref{sec:p-2p} for the general case after some straightforward modifications. This starts with an adjustment of the dimensions of the matrix $D_2$ to $(p+y)\times(p+y)$, with element (1,1) changed to $p-v$ rather than $p-2$, and all of the remaining terms adjusted accordingly. The dimensions of the vectors $U, B$ and $N$ are also changed to $p+y$, with $U$ and $N$ potentially containing additional nonzero elements. We also redefine the analytic solution $U$ as
\begin{align}
    U^T=\begin{bmatrix}
A_1& 
A_2& 
\hdots & A_{y}&
0 &
\hdots &
    \end{bmatrix}
\end{align}
for a vector of length $p+y$ containing $y$ nonzero elements.  

\begin{table}[h!]
\centering
\begin{center}
\hspace*{-1cm}\begin{tabular}{|>{\centering\arraybackslash}p{7mm}|>{\centering\arraybackslash}p{7.8mm}|>{\centering\arraybackslash}p{7.5mm}|>{\centering\arraybackslash}p{7.5mm}|>{\centering\arraybackslash}p{7.5mm}|>{\centering\arraybackslash}p{0.75cm}|>{\centering\arraybackslash}p{0.75cm}|>{\centering\arraybackslash}p{0.75cm}|>{\centering\arraybackslash}p{1.3cm}|>{\centering\arraybackslash}p{1.9cm}|>{\centering\arraybackslash}p{0.75cm}|>{\centering\arraybackslash}p{1.4cm}|>{\centering\arraybackslash}p{0.95cm}|}
 \hline
 & \multicolumn{12}{|c|}{Power of Hyperbolic Secant terms } \\ & \multicolumn{12}{|c|}{generated by terms in the general NLS} \\
\hline
$\mu$ & $p-v$ & $\cdots$ & $p-4$ & $p-2$ & $p$ & & & & & & &\\
\hline
$\beta_2$ & $p-v$ & $\cdots$ & $p-4$& $p-2$ & $p$ & $p+2$ & & & & & & \\
\hline
$\beta_4$ & $p-v$ & $\cdots$ & $p-4$& $p-2$ & $p$ & $p+2$ & $p+4$ & & & & &\\
 \hline
 $\vdots$ & $\vdots$ & $\vdots$ & $\vdots$ & $\vdots$ & $\vdots$ & $\vdots$ & $\vdots$ & $\ddots$  &$\ddots$ &$\ddots$ & $\ddots$&\\
\hline
$\beta_n$ & $p-v$ & $\cdots$ & $p-4$ & $p-2$ & $p$ & $p+2$ & $\cdots$ & $3(p-v)$& $3(p-v)+2$ &$\cdots$& $p+n-2$& $p+n$\\
\hline
$\gamma$ & &  &  & & & & & $3(p-v)$& $3(p-v)+2$ &$\cdots$&$3p-2$&$3p$\\
\hline
\end{tabular}
\end{center}
\caption{The rows and columns of the table are grouped in the same way as Table~\ref{table:1} and \ref{table:2} for a substitution of Eq.~\eqref{eq:ppgen} into Eq.~\eqref{eq:stationaryeq}.}
\label{table:3}
\end{table}
In Sec.~\ref{sec:p-2p} the two lowest power columns were proportional and thus defined the ratio between amplitudes $A_{1,2}$. In Table~\ref{table:3}, the first $y$ columns are similarly proportional, and allow us to derive a relationship between all the amplitude terms. We follow the same approach as in Sec.~\ref{sec:analyticpp2}, where the ratio of prefactors for $\mu$ $(A_1/A_r)$ must be equal to the ratio of prefactors for $\beta_2$ ($A_1(p-v)^2/(A_r(p-2(y-r))^2-A_{r-1}(p-2(y-r)-2)(p-2(y-r)-1))$). This gives the general ratio between any two successive amplitudes
\begin{align}
\label{eq:genratio}
    \frac{A_r}{A_{r-1}}=\frac{(p-2(y-r)-2)(p-2(y-r)-1)}{(p-2(y-r))^2-(p-v)^2},
\end{align}
for arbitrary amplitude $A_r$. This allows us to find and inductively define each of the amplitudes in the general case, in terms of $A_1$. 

Then the nonlinear term $N$ requires us to cube an arbitrary number of terms, giving
\begin{align}\hspace*{-0.4cm}
    N^T=\gamma\begin{bmatrix}
\hdots\ 
0 &
A_1^3&
3 A_1^2A_2\ 
\hdots\ 6A_1A_2A_3\ \hdots\  6A_{y-2}A_{y-1}A_{y} \
\hdots \
3 A_{y-1}A_y^2 &
A_y^3
    \end{bmatrix}
\end{align}
where the prefactors of each term is determined by the multinomial coefficient. This results in $3y-2$ nonzero terms in the vector $N$, corresponding to hyperbolic secant terms between powers of $3(p-v)$  and $3p$. 

With all the matrices and vectors from Sec.~\ref{sec:p-2p} now known for the general case, we use Eq.~\eqref{eq:stationarymatrix} to follow the same process, generating a triangular matrix and solving a set of linear simultaneous equations. After choosing values for the two highest order dispersion terms ($\beta_n$ and $\beta_{n-2}$) and the nonlinear parameter ($\gamma$), we can obtain all remaining unknowns. Using the first $y$ columns, we have obtained a relationship between all the unknown amplitudes allowing us to write all of the amplitudes in terms of $A_1$, and similar to Sec.~\ref{sec:p-2p}, we then solve the two highest power linear equations simultaneously, to obtain values for $\alpha$ and $A_1$. Solving the remaining unused equations gives us the full dispersion relation and the nonlinear phase shift for Eq.~\eqref{eq:ppgen}. 

This method can be applied provided the nonlinear terms do not overlap with the set of linearly dependent columns at powers $p-v, \ldots, p$. This condition can be enforced by ensuring that the size of the elements $(p+y)$, is larger than or equal to the sum of equivalent columns $y$ and nonlinear terms $3y-2$, giving the general condition $p\geq3y-2$. Using this, we see when the power of the lowest hyperbolic secant term is $1$ ({\sl i.e.}, $p-v=1$), this inequality does not hold, consistent with our analysis in Sec.~\ref{sec:analyticpp2}.

\begin{figure}[hbt]
\centering
\hspace*{0cm}\includegraphics[width=130mm,clip = true]{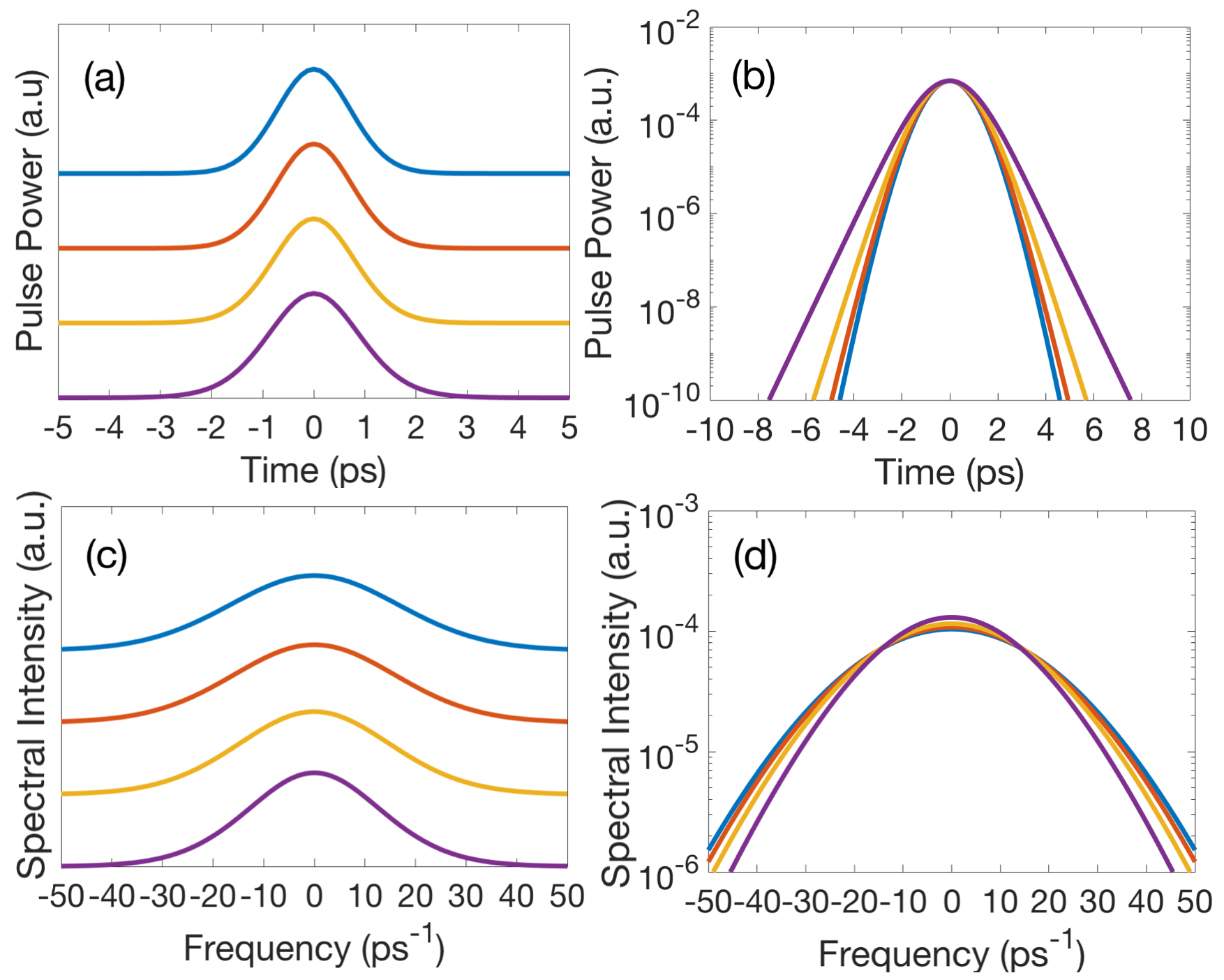}
\vskip 0mm
\caption{Solutions in the form Eq.~\eqref{eq:ppgen} for $n=20$ are plotted with $\alpha=0.3106~{\rm ps^{-1}}$, peak power $P = 0.6975~{\rm mW}$ and $\gamma = 1~{\rm W}^{-1} {\rm mm}^{-1}$ kept constant, for $v=0$ (blue curve: $\beta_{18}=-1\ {\rm ps^{18}~ mm}^{-1}$, $\beta_{20}=-1\ {\rm ps^{20}~ mm}^{-1}$, ${\rm FWHM} = 1.7056~{\rm ps}$), $v=2$ (red: $\beta_{18}=-0.5778\ {\rm ps^{18}~ mm}^{-1}$, $\beta_{20}=-0.4444\ {\rm ps^{20}~ mm}^{-1}$, ${\rm FWHM} = 1.7688~{\rm ps}$), $v=4$ (yellow: $\beta_{18}=-0.2490\ {\rm ps^{18}~ mm}^{-1}$, $\beta_{20}=-0.1624\ {\rm ps^{20}~ mm}^{-1}$, ${\rm FWHM} =   1.8782~{\rm ps}$), $v=6$, (purple: $\beta_{18}=-0.07277\ {\rm ps^{18}~ mm}^{-1}$, $\beta_{20}=-0.04281\ {\rm ps^{20}~ mm}^{-1}$, ${\rm FWHM} = 2.1015~{\rm ps}$) on (a) a linear scale, and (b) a logarithmic scale. Associated spectra on (c) a linear scale, and (d) a logarithmic. The solutions on the linear scale have been shifted for clarity.}
\label{fig:Solutionsgen}
\vskip-1mm
\end{figure}
\begin{figure}[hbt]
\centering
\hspace*{0cm}\includegraphics[width=100mm,clip = true]{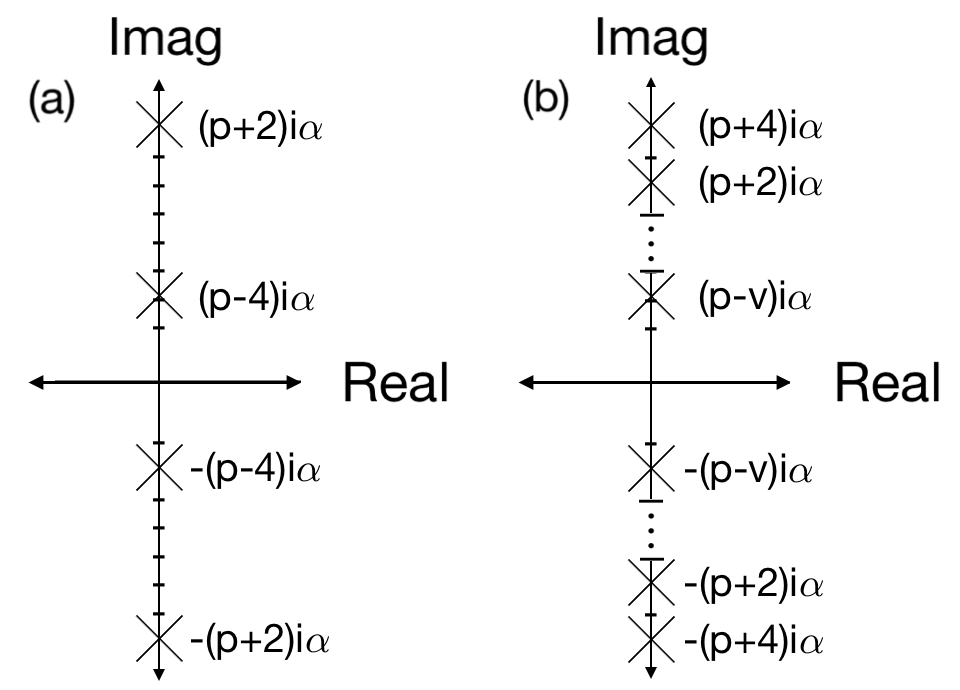}
\vskip 0mm
\caption{The positions of the poles for the Fourier Transform for (a) $A_1{\rm sech^{p-4}}(\alpha\tau) +$ $A_2{\rm sech^{p-2}}(\alpha\tau)$ $+ A_3{\rm sech^{p}}(\alpha\tau)$ and (b) $A_1{\rm sech^{p-v}}(\alpha\tau)+A_2{\rm sech^{p-v+2}}(\alpha\tau)+...$ $+ A_{y}{\rm sech^{p}}(\alpha\tau)$.}
\label{fig:polesgen}
\vskip-1mm
\end{figure}

For the general case, we can also modify the analytic form of $A$ and $\alpha$, which we found for the two term case in Eq.~\eqref{eq:Aexpp2} and Eq.~\eqref{eq:alphaexpp2}. This requires a simple modification of these equations by the elements dependent on the ratio between the amplitudes, and correcting these for the general case gives us the expression
\begin{align}
\alpha^2 = &\frac{n(n-1)}{\frac{n}{24}(13n^2-24n+8)-\frac{n-4}{8}\frac{A_{y-1}}{A_y}\left(2(n-2)-\frac{6(3n-4)(3n-2)}{n-4}\right)}~\frac{\beta_{n-2}}{\beta_n},\label{eq:alphaexgen}\\
\gamma A_1^2 &=-\frac{\beta_n\alpha^n}{3} \left(\frac{A_{1}}{A_y}\right)^2 \frac{(3n/2)!}{n!(n/2)!},\label{eq:Aexgen}
\end{align}
Here we normalised the ratio-dependent terms from Eq.~\eqref{eq:Aexpp2} and Eq.~\eqref{eq:alphaexpp2} and replaced these with the appropriate ratios we know for the general case. We find these analytic results to be consistent with the results we obtain numerically using the matrix method.

\subsection{Properties of the solutions}
In Fig.~\ref{fig:Solutionsgen}, we now plot four solutions with the same peak dispersion order $n=20$, using our generalised method to find the unknowns for the solutions. The principles of the tail argument outlined in Sec.~\ref{sec:analytic} once again apply, where the smallest real part of the tail solution dominates. Since we expect the solution of Eq.~\eqref{eq:ppgen} to have a dominant term at $e^{\pm (p-v)\alpha\tau}$, we expect the steepest tail of these solutions to be for $v=0$, with an increasing value of $v$ corresponding to decreasing steepness, which we observe clearly in Fig.~\ref{fig:Solutionsgen}(b). The solution in the spectral domain in Fig.~\ref{fig:Solutionsgen}(d) shows an opposite effect, where the steeper solutions in the temporal domain correspond to shallower solutions in the spectral domain, which is what we would expect.

We also now analyse the Fourier transform for the general case, where a substitution of the appropriate amplitude using Eq.~\eqref{eq:genratio}, allows us to rearrange the general Fourier transform into the form
\begin{align}
    &{\rm FT}\left(A_1\ {\rm sech^{p-v}}(\alpha\tau)+A_2\ {\rm sech^{p-v+2}}(\alpha\tau)+...+A_{y}\ {\rm sech^{p}}(\alpha\tau)\right)=\nonumber \\
    &\frac{A_1}{(4\alpha^2)^{v/2}}\frac{(\omega^2+\alpha^2(p-v+2)^2)(\omega^2+\alpha^2(p-v)^2)...(\omega^2+\alpha^2p^2)}{(p-v+1)(p-v+3)...(p-1)}~{\rm FT}({\rm sech^{p-v}}(\alpha\tau)).
    \label{eq:FTgenorders}
\end{align}
We recall from Sec.~\ref{sec:prop}, that for the Fourier transform of the basic hyperbolic secant of power one, there are simple poles at $\omega = \pm i\alpha, \pm 3i\alpha, \pm 5i\alpha, ...$, however for an arbitrary hyperbolic secant at power $r$, all simple poles below this power are supplanted. We then observed in Sec.~\ref{sec:prop2} that for an additional hyperbolic secant term in the temporal solution, the Fourier transform had an additional simple pole supplanted at $\omega = \pm pi\alpha$, producing a gap in the simple pole structure. For the Fourier transform of the general case, we now expect simple poles at $\omega = \pm (p-v)i\alpha, \pm (p-v+2)i\alpha, \pm (p-v+4)i\alpha, ...$, but now with additional poles supplanted at 
$\omega = \pm (p-v+2)i\alpha, ..., \pm pi\alpha$, 
due to the amplitude ratios (Eq.~\eqref{eq:genratio}) from the additional roots generated by the temporal terms. 
We see this pole structure shown in Fig.~\ref{fig:polesgen}, and note that the change in the analytic solution in the temporal domain again corresponds to a clear change in the structure of the Fourier transform, with an increasing number of simple poles supplanted corresponding to an increasing number of terms in Eq.~\eqref{eq:ppgen}. In fact, knowing the structure of the Fourier transform provides a method of identifying the number and power of the terms for the solution in the temporal domain. We discuss this method further in Sec.~\ref{sec:disc}.

\section{Discussion and Conclusion}\label{sec:disc}


The results we have presented provide an analytical framework for exact solutions for higher order dispersion systems with anomalous dispersion, and have the potential to be expanded upon both analytically and experimentally. Recent studies have shown the capability of fibre lasers with programmable dispersion demonstrating soliton solutions up to $10^{\rm th}$ order dispersion \cite{Runge_2020, runge_qiang_alexander_rafat_hudson_blanco_redondo_desterke_2021b}, so experimental observation of the analytic solutions found in this paper is realistic. Further study of their properties may help to identify applications. 

These analytic solutions can be used to improve the numerical methods used to find additional solitons as solutions of high-order differential equations. An effective method to find solutions of Eq.~\eqref{eq:stationaryeq} is the Newton Conjugate-Gradient method \cite{Yang_2009, Yang_2010}. It ideally requires an exact soliton solution at the same order of dispersion, corresponding to the same order of the differential equation, as the desired one. The solutions found in this work can therefore serve as starting points for the numerical method. This then allows us to move adiabatically through solution space without changing the order of the differential equation until the solution at the desired parameters is found. Observing the rate of convergence also allows us to examine the effectiveness of this numerical method for higher order dispersion. 

We have presented results of Eq.~\eqref{eq:stationaryeq}, which is a stationary equation, but this does not mean that these solutions are stable. Determining stability requires the Eq.~\eqref{eq:gnls}. We carried out simulations of the full evolution of the solutions we have presented using the split step Fourier method \cite{Agrawal_2013}. Although not carried systematically, all of the solutions we have tested, which include up to three terms and dispersion up to $14^{\rm th}$ order, appear to be stable. However, a full stability analysis is outside the scope of this paper. 

A consequence of the pole structure of the multi-term solutions, and the removal of some of the poles, is that the asymptotic expansion of the solutions in Eq.~\eqref{eq:ppgen} can be written as  
\begin{equation}\label{eq:asymp}
   B_{1}~ e^{\pm(p-v)\alpha\tau} + B_{y+1}~ e^{\pm(p+2)\alpha\tau} + B_{y+2}~ e^{\pm(p+4)\alpha\tau}  \ldots,
\end{equation}
where the $B_i$ are linear combinations of the $A_i$ from Section~\ref{sec:allp}, and the terms $B_2,\ldots,B_y$ vanish. In other words, the number of missing terms in the expansion equals the number of poles that have been removed. One may conclude that as $v$ becomes very large, then the solution increasingly resembles the function $e^{-(p-v)\alpha|\tau|}$; however, since this function has a discontinuous derivative at the origin it is unlikely that it can ever be reached. Nonetheless, result~\eqref{eq:genratio} for the ratio of amplitudes of consecutive terms in the expression for the soliton can also be found by requiring that the $2^{\rm nd}$ derivative of Eq.~\eqref{eq:ppgen} is proportional to the equation itself. Thus, even though the positions of the simple poles are easy to predict, the implications of some of the simple poles being supplanted and the significance of a reduced number of simple poles as dispersion order increases, remain open questions. 

The analytic solutions we construct consist of hyperbolic secant terms with powers in successive multiples of two. For solutions including successive terms with a larger difference than two, such as 
\begin{align}
    u = ...+A_{r-d}\ {\rm sech}^{c}(\alpha\tau)+A_{r}\ {\rm sech}^{c+2d}(\alpha\tau)+...,
    \label{eq:sechgap}
\end{align}
for cases where $d>1$, we find that the method applied in this paper is not applicable. The reason for this is simple. In Sec.~\ref{sec:allp} we found the ratio of successive amplitudes; for a solution with a gap between successive powers larger than two, $A_{r-1}=0$ is enforced, for which the only solution is $r=1$, which prohibits solutions with more than one hyperbolic secant term. Thus we cannot easily find the ratio between amplitudes, preventing us from applying the method we developed.

In the temporal domain, we exploited the root structure for the single term solutions. However for the solutions with multiple hyperbolic secant terms, some of the characteristic tail roots remain analytically unpredictable due to the presence of nonlinear terms. Numerical calculations shows that all roots remain on the real axis for the two-term solitons, with complex roots introduced for solutions with more terms, 
and that the roots linearly scale with $\alpha$. Being able to analytically predict the tail root structure for an arbitrary solution would allow us to apply the same method as in Sec.~\ref{sec:analytic} to find the associated dispersion relation, which would be even more straightforward than our current method. 

In summary, we present a study of exact analytic stationary solutions consisting of a superposition of hyperbolic secant terms of different integer powers, and a corresponding method to find the associated dispersion relations. In Sec.~\ref{sec:analytic}, we considered solutions with a single term, and derived their form for arbitrary dispersion order using a constructed polynomial to define the dispersion relation. Then in Sec.~\ref{sec:analyticpp2}, we find the form of solutions with two terms for arbitrary dispersion order, and present a convenient method using to find the associated dispersion relation, and in Sec.~\ref{sec:allp}, generalise these results to an arbitrary superposition of terms. These studies all include a consideration of the distinct pole structure for any unique solution in the spectral domain.

\ack
We acknowledge funding from the Australian Research Project (ARC) Discovery Project (DP180102234) and the Asian Office of Aerospace R\&D (AOARD) grant (FA2386-19-1-4067).

\appendix
\section{Analytic Solution: $Sech+Sech^3$}
\label{sec:sech13}

\begin{figure}[hbt]
\centering
\hspace*{0cm}\includegraphics[width=120mm,clip = true]{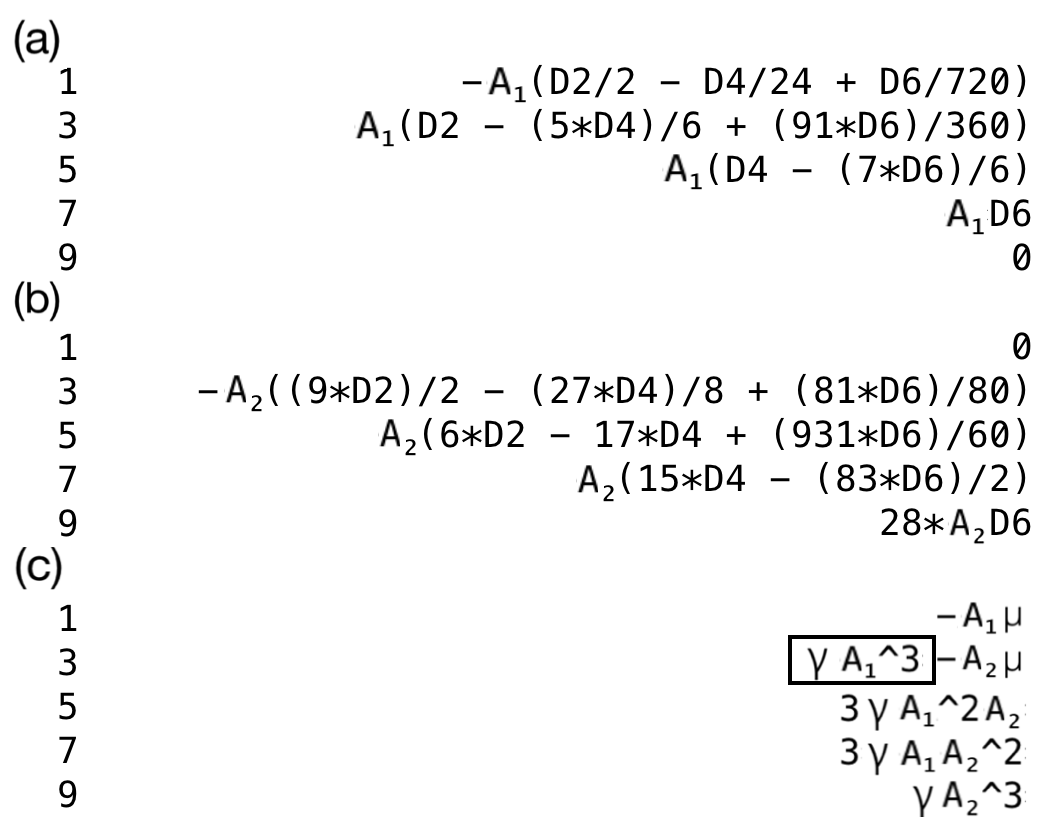}
\vskip 0mm
\caption{\label{fig:p3ex}
Terms at each hyperbolic secant power when Eq.~\eqref{eq:sech3sech} is substituted into Eq.~\eqref{eq:stationaryeq}. (a) terms generated by dispersion applied to $A_1\ {\rm sech}(\alpha\tau)$, and (b) $A_2\ {\rm sech}^{3}(\alpha\tau)$, and (c) due to $\mu u$ and $\gamma u^3$. The black box shows the nonlinear term at power $3$. $D(r)\equiv\beta_{r}\alpha^r$. }
\vskip-1mm
\end{figure}

Here we consider a solution, also found by Kudryashov \cite{Kudryashov_2020}, where the systematic approach we present in Sec.~\ref{sec:p-2p} cannot be applied, and so the solution and the associated dispersion relation cannot be separated.
We consider the analytic solution of the form
\begin{align}
    u = A_1\ {\rm sech(\alpha\tau)}+A_2\ {\rm sech^{3}}(\alpha\tau).
    \label{eq:sech3sech}
\end{align}
In Sec.~\ref{sec:analyticpp2} we argue that for $p<4$, the linear algebra methods described do not apply, because the columns at power~1 and~3 are not proportional, due to a nonlinear term at power 3 (see Fig.~\ref{fig:p3ex}(c)). This prevents us from using the ratio Eq.~\eqref{eq:AB} to find a relationship between $A_1$ and $A_2$, leaving both as unknowns. Finding this analytic solution requires us to simultaneously solve all five equations without the simplifications from Sec.~\ref{sec:analyticpp2}. The terms at each power which the equations consist of are grouped by hyperbolic secant power in Fig.~\ref{fig:p3ex}, for Eq.~\eqref{eq:sech3sech} substituted into Eq.~\eqref{eq:stationaryeq}. 

We aim to make a series of substitutions between these equations, in order to isolate the relationship between the factors $A_1$ and $A_2$. We achieve this by eliminating all $\beta_{2,4,6}$ and $\alpha$ terms. We begin by obtaining an expression for $\mu$ from the equation at power 1, and substituting this into the equation at power 3. Following this, we make a series of substitutions of the equation at power 9, into the equations at power 7 and 5. This allows us to write the dispersion parameters, and $\alpha$, in terms of $\gamma$ and $A_1$ and $A_2$, giving
\begin{equation}
    \beta_{4}\alpha^{4}=-\frac{83\gamma}{840}(2A_1A_2+A_2^{2}),
    \quad 
    \beta_{2}\alpha^{2}=-\frac{\gamma}{2520}\!\left(\!1177A_1^{2}+1387A_1A_2+\frac{1891A_2^{2}}{4}\right)\!.\label{eq:b24alpha}
\end{equation}
Substituting these two equations and our equation at power~9 into our updated equation at power~3, gives
\begin{equation}
    \gamma(1343A_1^{3}+3736A_1^{2}A_2+3600A_1A_2^{2}+1152A_2^{3})=0
    \label{eq:ABfinal}
\end{equation}
dividing through by the nonzero factor $\gamma$ we are left with a cubic equation composed of only the parameters $A_{1,2}$. Solving these equations gives $A_1=-\frac{12}{17} A_2$, consistent with Kudryashov \cite{Kudryashov_2020}. Substitutions of this ratio into prior equations allows us to obtain relationships between all the parameters, giving the parameters in Table~\ref{table:terms2}. 


\begin{table}[h!]
\renewcommand{\arraystretch}{1.5}
\centering
\begin{center}
\begin{tabular}{|>{\centering\arraybackslash}p{11mm}|>{\centering\arraybackslash}p{70mm}|}
\hline
$\alpha^2$ & $\frac{510}{581}\frac{\beta_{4}}{|\beta_{6}|}$\\
\hline
$\gamma A_2^2$ & $28\alpha^{6}|\beta_{6}|$\\
\hline
$\mu$ & $\frac{804}{1627}\alpha^6|\beta_{6}|$ \\
\hline
$\beta_2$ & $-\frac{92659}{104040}\alpha^4|\beta_{6}|$\\
\hline
\end{tabular}

\caption{Exact expressions for the unknowns of Eq.~\eqref{eq:sech3sech}}
\label{table:terms2}
\end{center}
\end{table}

The method used to obtain the unknowns in Eq.~\eqref{eq:sech3sech} are significantly more demanding than solutions for $p\geq4$, and we predict that solutions in the general case from Sec.~\ref{sec:allp} for the condition $p<3y-2$ will similarly be practically demanding to solve.

By the same argument as for the exact solution in Sec. \ref{sec:p=p}, we consider the asymptotic expansion of Eq. \eqref{eq:sech3sech}, and similarly expect a superposition of terms of the type
$e^{\pm\alpha\tau}$,
$e^{\pm3\alpha\tau}$, $e^{\pm5\alpha\tau}$, etc. We note that in this case however, the $e^{\pm\alpha\tau}$ term is the only term generated linearly, as all other terms can be generated from the third harmonic of Eq. \eqref{eq:sech3sech}. This results in a highly unique root structure, where applying Eq.~\eqref{eq:character} produces two real roots at $\pm p\alpha$ and a quartet of complex roots, which have a smaller real part than the real roots. Nonetheless, the tails are dominated by the real roots. 
The pole structure found for solutions in Sec.~\ref{sec:prop} is also absent.

\section*{References}
\bibliographystyle{iopart-num}
\bibliography{sol.bib}

\providecommand{\noopsort}[1]{}\providecommand{\singleletter}[1]{#1}%
\providecommand{\newblock}{}
\begin{thebibliography}{10}
\expandafter\ifx\csname url\endcsname\relax
  \def\url#1{{\tt #1}}\fi
\expandafter\ifx\csname urlprefix\endcsname\relax\def\urlprefix{URL }\fi
\providecommand{\eprint}[2][]{\url{#2}}

\bibitem{Scott_Chu_McLaughlin_1973}
Scott A, Chu F and McLaughlin D 1973 {\em Proceedings of the IEEE\/} {\bf 61}
  1443–1483

\bibitem{Polturak_1981}
Polturak E, deVegvar P~G~N, Zeise E~K and Lee D~M 1981 {\em Phys. Rev. Lett.\/}
  {\bf 46} 1588--1591

\bibitem{r}
Akhmediev N~N and Ankiewicz A 2011 {\em Dissipative solitons: from optics to
  biology and medicine\/} (Springer) ISBN 9783642062391

\bibitem{Nakazawa_1994}
Nakazawa M 1994 {\em IEEE Communications Magazine\/} {\bf 32} 34–41

\bibitem{Mollenauer_1991}
Mollenauer L~F, Neubelt M~J, Haner M, Lichtman E, Evangelides S~G and Nyman B~M
  1991 {\em Electronic Lett.\/} {\bf 27} 2055--2056

\bibitem{Haus_1996}
Haus H~A and Wong W~S 1996 {\em Rev. Mod. Phys.\/} {\bf 68} 423--444

\bibitem{Dudley_2006}
Dudley J~M, Genty G and Coen S 2006 {\em Reviews of Modern Physics\/} {\bf 78}
  1135–1184

\bibitem{n}
Husakou A~V and Herrmann J 2001 {\em Physical Review Letters\/} {\bf 87}

\bibitem{o}
Mitschke F~M and Mollenauer L~F 1987 {\em Optics Letters\/} {\bf 12} 407

\bibitem{Turitsyn_2012}
Turitsyn S~K, Bale B~G and Fedoruk M~P 2012 {\em Physics Reports\/} {\bf 521}
  135--203

\bibitem{Elgin_1992}
Elgin J~N 1992 {\em Opt. Lett.\/} {\bf 17} 1409--1410

\bibitem{Kodama_1994}
Kodama Y, Romagnoli M, Wabnitz S and Midrio M 1994 {\em Opt. Lett.\/} {\bf 19}
  165--167

\bibitem{Hook_1993}
Höök A and Karlsson M 1993 {\em Opt. Lett.\/} {\bf 18} 1388--1390

\bibitem{Aceves_1994}
Aceves A~B, Angelis C~D, Nalesso G and Santagiustina M 1994 {\em Opt. Lett.\/}
  {\bf 19} 2104--2106

\bibitem{Blanco_Redondo_2016}
Blanco-Redondo A, de~Sterke C~M, Sipe J~E, Krauss T~F, Eggleton B~J and Husko C
  2016 {\em Nature Comm.\/} {\bf 7} 10427

\bibitem{Runge_2020}
Runge A~F~J, Hudson D~D, Tam K~K~K, de~Sterke C~M and Blanco-Redondo A 2020
  {\em Nat. Photonics\/} {\bf 14} 492–497

\bibitem{runge_qiang_alexander_rafat_hudson_blanco_redondo_desterke_2021b}
Runge A~F~J, Qiang Y~L, Alexander T~J, Rafat M~Z, Hudson D~D, Blanco-Redondo A
  and de~Sterke C~M 2021 {\em Physical Review Research\/} {\bf 3}

\bibitem{deSterke_2021}
de~Sterke C~M, Runge A~F~J, Hudson D~D and Blanco-Redondo A 2021 {\em APL
  Photonics\/} {\bf 6} 091101 (\textit{Preprint}
  \eprint{https://doi.org/10.1063/5.0059525})
  \urlprefix\url{https://doi.org/10.1063/5.0059525}

\bibitem{Karlsson_1994}
Karlsson M and Höök A 1994 {\em Opt. Commun.\/} {\bf 104} 303--307

\bibitem{Piche_1996}
Pich\'e M, Cormier J~F and Zhu X 1996 {\em Opt. Lett.\/} {\bf 21} 845--847

\bibitem{Qiang_2022}
Qiang Y~L, Alexander T~J and de~Sterke C~M 2022 {\em Phys. Rev. A\/} {\bf
  105}(2) 023501
  \urlprefix\url{https://link.aps.org/doi/10.1103/PhysRevA.105.023501}

\bibitem{Kudryashov_2020}
Kudryashov N~A 2020 {\em Applied Mathematics and Computation\/} {\bf 371}
  124972

\bibitem{hosseini2021}
Hosseini K, Sadri K, Mirzazadeh M, Chu Y, Ahmadian A, Pansera B and Salahshour
  S 2021 {\em Results in Physics\/} {\bf 23} 104035

\bibitem{kudryashov_2021a}
Kudryashov N~A 2021 {\em Optik\/} {\bf 235} 166626

\bibitem{kudryashov_2021b}
Kudryashov N~A 2021 {\em Mathematics\/} {\bf 9} 3024

\bibitem{arnous2022}
Arnous A~H, Zhou Q, Biswas A, Guggilla P, Khan S, Yıldırım Y, Alshomrani A~S
  and Alshehri H~M 2022 {\em Physics Letters A\/} {\bf 422} 127797

\bibitem{Zakharov_1972}
Zakharov V and Shabat A~B 1972 {\em Soviet Journal of Experimental and
  Theoretical Physics\/} {\bf 34} 62

\bibitem{Ablowitz_Segur_1981}
Ablowitz M~J and Segur H 1981 {\em Solitons and the inverse scattering
  transform\/} (Society For Industrial And Applied Mechanics) ISBN
  9780898714777

\bibitem{erdelyi_bateman_1954}
Erdélyi A and Bateman H 1954 {\em Tables of integral transforms. vol. 1.\/}
  (New York)

\bibitem{Yang_2009}
Yang J 2009 {\em J. Comput. Phys.\/} {\bf 228} 7007--7024

\bibitem{Yang_2010}
Yang J 2010 {\em Nonlinear waves in integrable and nonintegrable systems\/}
  (Society For Industrial And Applied Mathematics) ISBN 9780898717051

\bibitem{Agrawal_2013}
Agrawal G~P 2013 {\em Nonlinear fiber optics\/} (Academic Press) ISBN
  9780123970237

\end{thebibliography}

\end{document}